\documentclass{aa}

\usepackage{amsmath}
\usepackage{graphicx}
\usepackage{txfonts}
\usepackage{url}
\usepackage{natbib}

\newcommand{\diag}{SPOC diagram} 
\newcommand{\diags}{SPOC diagrams} 

\newcommand{\degree}{\ensuremath{^\circ}}
\newcommand{\Tl}{\textit{$T$}} 
\newcommand{\tpl}{\textit{$t_p$}} 
\newcommand{\al}{\textit{$a$}} 
\newcommand{\el}{\textit{$e$}} 
\newcommand{\il}{\textit{$i$}} 
\newcommand{\wl}{\textit{$\omega_{p}$}} 
\newcommand{\Ol}{\textit{$\Omega$}} 
\newcommand{\nl}{\textit{$\nu$}} 
\newcommand{\frl}{\textit{$f_T$}} 
\newcommand{\rl}{\textit{$r_{orbit}$}} 
\newcommand{\pal}{\textit{$\alpha$}} 
\newcommand{\sepl}{\textit{$\theta$}} 

\newcommand{\lol}{\textit{$\lambda_{Obs}$}} 

\newcommand{\sal}{\textit{$w_R$}} 
\newcommand{\Agl}{\textit{$A_{g}$}} 
\newcommand{\Agrl}{\textit{$A_{g,Rayleigh}$}} 
\newcommand{\Agtl}{\textit{$A_{g,Total}$}} 
\newcommand{\Agll}{\textit{$A_{g,Lambertian}$}} 
\newcommand{\phl}{\textit{$A_{\alpha}$}} 
\newcommand{\pfl}{\textit{$\phi$}} 

\newcommand{\dpl}{\textit{$Pz$}} 
\newcommand{\apl}{\textit{$\Phi_{Pz}$}} 
\newcommand{\crl}{\textit{$CR$}} 

\bibpunct{(}{)}{;}{a}{}{,}

\begin{document}

\title{Predicting exoplanet observability in time, contrast, separation and polarization, in scattered light}
\titlerunning{Predicting exoplanet observability in contrast, separation and polarization}

   \author{Guillaume Schworer\inst{1,2} \& Peter G. Tuthill\inst{2}}

   \institute{LESIA, Observatoire de Paris, CNRS/UMR 8109, UPMC, Université Paris Diderot, 5 place J. Janssen, 92195 Meudon, France
              \email{guillaume.schworer@obspm.fr}
         \and
             Sydney Institute for Astronomy (SIfA), School of Physics, The University of Sydney, NSW 2006, Australia}

   \date{Received January 1, 2014; accepted January 1, 2014}

  \abstract
   {Polarimetry is one of the keys to enhanced direct imaging of exoplanets. Not only does it deliver a differential observable providing extra contrast, but when coupled with spectroscopy, it also reveals valuable information on the exoplanetary atmospheric composition. Nevertheless, angular separation and contrast ratio to the host-star make for extremely challenging observation. Producing detailed predictions for exactly how the expected signals should appear is of critical importance for the designs and observational strategies of tomorrow's telescopes.}
   {We aim at accurately determining the magnitudes and evolution of the main observational signatures for imaging an exoplanet: separation, contrast ratio to the host-star and polarization as a function of the orbital geometry and the reflectance parameters of the exoplanet.}
   {These parameters were used to construct polarized-reflectance model based on the input of orbital parameters and two albedo values. The model is able to calculate a variety of observational predictions for exoplanets at any orbital time.}
   {The inter-dependency of the three main observational criteria -- angular separation, contrast ratio, polarization -- result in a complex time-evolution of the system. They greatly affect the viability of planet observation by direct imaging. We introduce a new generic display of the main observational criteria, which enables an observer to determine whether an exoplanet is within detection limits: the Separation-POlarization-Contrast diagrams (SPOC).}
   {We explore the complex effect of orbital and albedo parameters on the visibility of an exoplanet. The code we developed is available for public use and collaborative improvement on the python package index, together with its documentation. It is another step towards a full comprehensive simulation tool for predicting and interpreting the results of future observational exoplanetary discovery campaigns.}


   \maketitle


\section{Introduction}
Observational technologies for detecting the light reflected from an exoplanet are reaching a level of precision that makes direct imaging of exoplanets a realistic possibility in about the coming decade. Because of the very challenging contrast ratio between exoplanets and host-stars in the optical, there is great interest in differential methods, such as polarimetry, to deliver an extended reach to imaging instruments (\citet{2006PASP..118.1302H},~\citet{2006SPIE.6269E..0TK},~\citet{2006dies.conf..165S}). Built upon previous detections of polarized signals from exoplanets  ~\citet{2008ApJ...673L..83B}, ~\citet{2009ApJ...696.1116W}, ~\citet{2009MNRAS.393..229L}, ~\citet{2011ApJ...728L...6B}, the polarimetric imaging modules that are currently being integrated on the 10m class telescopes (SPHERE-ZIMPOL on the VLT~\citet{2008SPIE.7014E..3FT}, GPI on Gemini~\citet{2012SPIE.8446E..91W}, and VAMPIRES on SUBARU Telscope~\citet{2015MNRAS.447.2894N}) promise to achieve a differential polarized contrast down to $\
approx 10^{-6}$ 
in the visible or near-
infrared. It is therefore an opportune moment to produce detailed predictions for exactly how the expected signals should appear, which will be of critical importance for the designs and observational strategies of these instruments. We aim at accurately determining the magnitudes and evolution of the main observational signatures as a function of the basic parameters of the exoplanetary system: the star-exoplanet orbital parameters and the optical properties of the planet. We incorporate the complexity arising from effects such as polarization based on Rayleigh scattering. Additionally and as a first-order observability estimator, the absolute flux in photon per unit time and surface received from the exoplanet target is computed assuming a black-bodied star. We finally provide a querier and parser of the~\url{http://exoplanet.eu/} exoplanet database for searching and importing any star-planet target.

Several models were already developed to predict the light signature of an exoplanet as it should appear to an observer, for example see~\citet{2010ApJ...724..189C}, ~\citet{2009A&A...504..259B}, or for the case of Earth-like planets with varying atmospheric parameters, see \citet{2012A&A...546A..56K} and ~\citet{2010ApJ...723.1168Z}. However, to the best of our knowledge, the inter-dependency of the main observational criteria is never taken carefully into account. The relative evolution of angular separation, polarization, and contrast ratio is of critical importance because their respective maxima do not occur at the same time in the general case: this makes the best-case scenario for detection very unlikely. We correct for this by showing the relative evolution of the angular separation, contrast ratio, and polarization as a function of the orbital and reflectance parameters of the planet.
The findings resulting from this integrated treatment highlight dependencies that are much more complex than previously reported. The three main observational criteria -- angular separation, contrast ratio, and polarization -- in general do not exhibit a strong positive correlation and must be analyzed separately to determine the direct visibility of a potential planet target.


\section{Model}
We developed a polarized reflectance model. Its algorithm relies on the orbital parameters of the exoplanet and the linear combination of two albedo values; it is briefly described below.

As our prime observables are in any event differential between the star and exoplanet (such as the contrast ratio) or inherent to the planet (such as the polarization fraction), absolute fluxes are not described in detail. A simple black-body modeling of the host stars is used to give a first-order idea on the absolute fluxes that are expected from a exoplanet. For the sake of conciseness, ``polarization'' is used instead of ``polarization fraction''.


\subsection{Orbital parameters}

Each exoplanet orbit is defined by the usual parameters listed in Table~\ref{orb_param}. Note that the shape of the orbit as seen by the observer is solely defined by \el, \il, and \wl.

\begin{table}[h!]
\centering
\begin{tabular}{ | l | l | l | }
\hline
  Symbol & Name & Unit \\ \hline\hline
  $\Tl$ & Period & days  \\ \hline
  $\tpl$ & Time at periapsis & Julian date \\ \hline
  $\al$ & Semi-major axis & AU \\ \hline
  $\el$ & Eccentricity & $\varnothing$ \\ \hline
  $\il$ & Inclination to the observer & degrees \\ \hline
  $\wl$ & Argument of periapsis & degrees \\ \hline
  $\Ol$ & Longitude of ascending node & degrees \\ \hline
\end{tabular}
  \caption{Orbital parameters for an exoplanet. \il=90$\degree$ corresponds to an edge-on orbit (transiting), \il>90$\degree$ corresponds to a retrograde orbit.}
  \label{orb_param}
\end{table}

The phase angle of the exoplanet and the distance between the star and the planet are obtained from these orbital parameters. The phase angle \pal is defined as the vector angle between the star, an exoplanet, and the observer, as seen from the exoplanet. It is $0\degree$ when the planet is at full phase (superior conjunction) and $180\degree$ when the exoplanet is at new phase (inferior conjunction, transiting).\par


\subsection{Rayleigh scattering polarization}

For the sake of conciseness, in the following discussions we use the quantity called phased albedo, which is defined as the product of the phase function and the geometric albedo of a planet, for a given wavelength:
  \begin{equation}\label{phasedalbedo}
      \phl(\pal)=\Agl \cdot \pfl(\pal)=\frac{I(\pal)}{\pi F},
  \end{equation}
for an incident flux $\pi F$, with $I(\pal)$ the emerging flux from a body at the phase angle \pal. It represents the fraction of the incident irradiance that is reflected by the planet when it is seen at phase angle \pal by the observer, so that $E_{Planet,~out}=\phl(\pal) \cdot E_{Planet,~in}$, with $E$ being the spectral irradiance arriving at or leaving the planet.

The polarization induced by an exoplanet on a reflected beam of light is described by many different models, such as Mie, Rayleigh, or Raman scattering. It is assumed here that single Rayleigh scattering is the predominant source of polarization in planetary atmospheres. This is a good approximation as long as Mie scattering (and especially the primary rainbow polarizing effect of liquid droplets in clouds, see~\citet{2007AsBio...7..320B} remains lower than Rayleigh scattering. This scenario is usually achieved for wavelengths shorter than approximately $1\mu m$, which also corresponds to the most favorable wavelengths for Rayleigh polarization measurements of scattered light, as it is function of $\lambda^{-4}$. Integrated over the stellar disk, the flux from the star can be considered to be unpolarized (\citet{1987Natur.326..270K}. Hence, only the light from the exoplanet carries polarization.

\begin{figure}
  \centering
  \includegraphics[width=\hsize]{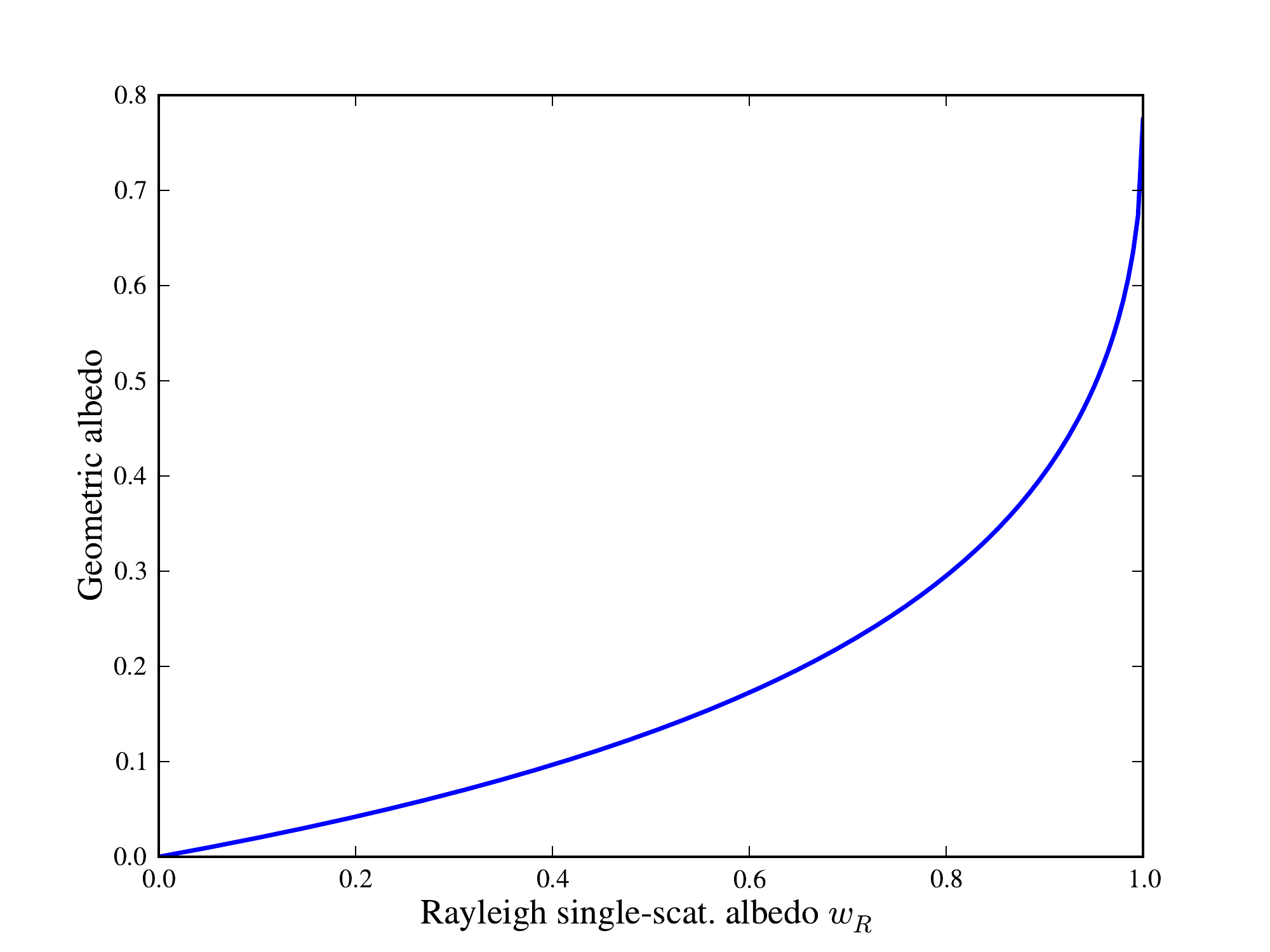}
    \caption{The geometric albedo \Agrl arising from single-scattering as a function of the Rayleigh single-scattering albedo \sal.}
  \label{geoalbfromscatfig}
  \end{figure}

The results of~\citet{2012ApJ...747...25M} for the Stokes parameters $Q/I$ and $U/I$ were adopted as input data to carry out more polarization and reflectance computations. Their model assumes an unresolved semi-infinite homogeneous atmosphere (hence cloud free), dominated by Rayleigh scattering. The atmosphere is assumed to be spherical, which is a satisfactory assumption when the planetary rotation remains slow. Given the strong depolarizing effect of multiple scattering, they used the single-scattering albedo as a unique reflectance parameter from which a geometric albedo \Agrl arising from the Rayleigh single-scattering albedo was calculated; their result is reproduced in Figure~\ref{geoalbfromscatfig}.

\begin{figure}[h!]
  \centering
  \includegraphics[width=\hsize]{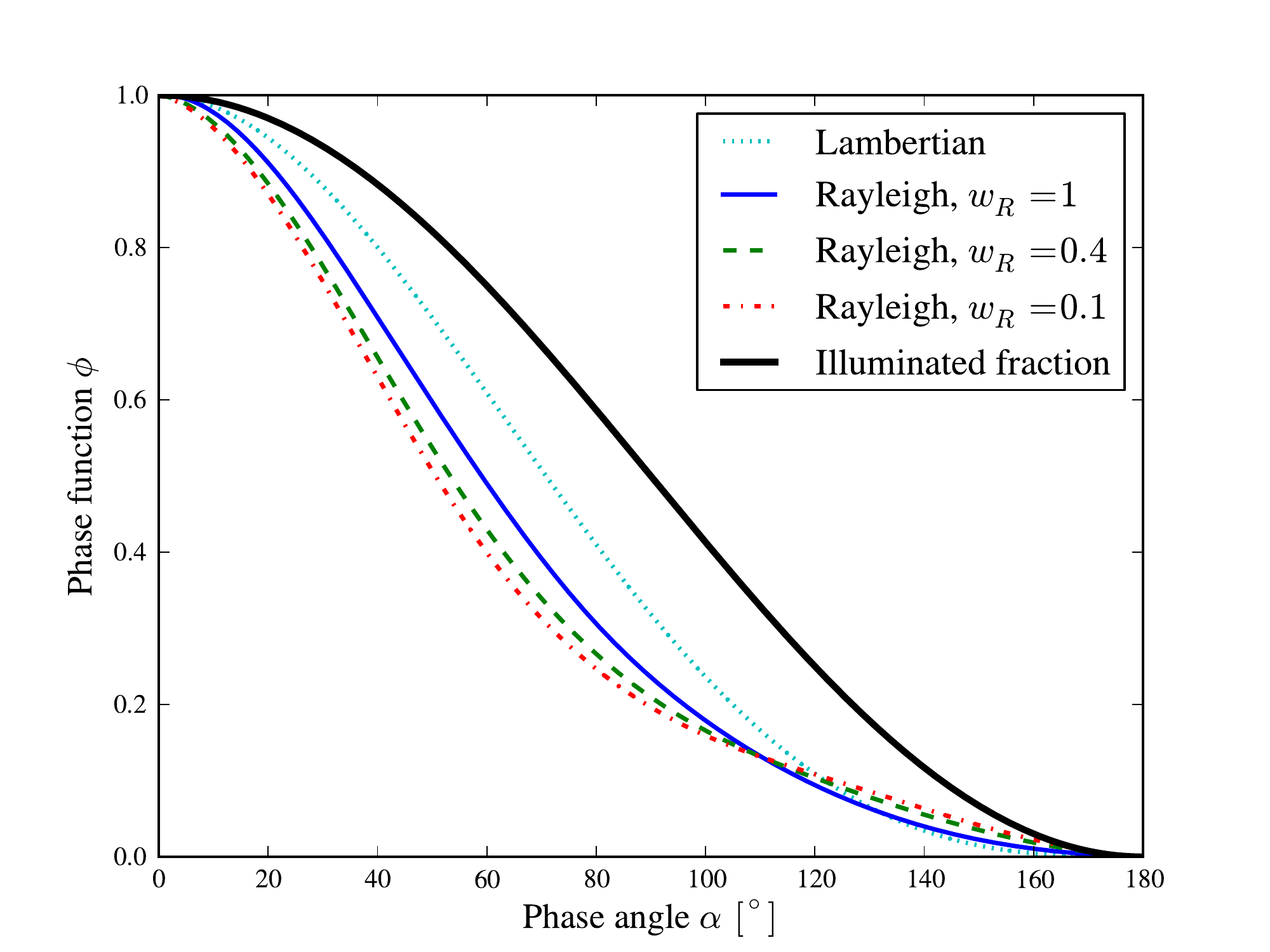}
    \caption{Phase functions \pfl as a function of phase angle \pal for several scattering models. The illuminated fraction is also displayed (thick line). The Rayleigh-scattering model corresponds to a vectorial Rayleigh phase matrix.}
  \label{phasefunctions}
  \end{figure}

Stokes-V is found to be zero for Rayleigh scattering in planet atmospheres. Hence, the degree of polarization $\dpl_{Rayleigh}$ is defined by
  \begin{equation}\label{polfractioneq}
      \dpl_{Rayleigh}=\frac{\sqrt{Q_{out}^2+U_{out}^2}}{I_{out}},
  \end{equation}

and it depends on the phase angle \pal of the planet and the single-scattering albedo \sal, which represents the amount of absorption versus scattering in a given atmosphere. The existing literature provides solutions to several phase functions.~\citet{2012ApJ...747...25M} provided an analytical solution for scattering models for Lambertian, Rayleigh, isotropic, and asymmetric scattering. \citet{2010ApJ...723.1168Z} addressed the liquid surface scattering model. Some phase functions are reproduced in Figure~\ref{phasefunctions}. Given the symmetry of the phase angle \pal along the orbit, only $\pal=[0,180]\degree$ values are represented. The illuminated fraction values are also plotted for comparison.

Rayleigh and Lambertian surface models follow the same general S-curve shape between zero flux and full flux. A Lambertian or Rayleigh planet is faint at small \pal because the illuminated fraction $f_I$ is small, hence fewer photons are reflected.


\subsection{Polarized and unpolarized albedo}

It is now apparent that clouds and condensates are very common in planet atmospheres. An atmosphere solely described by a single-scattering albedo could only take into account Rayleigh scattering of the atmosphere in the optically thick case, which is quite restrictive. Indeed, clouds, rocky surfaces or liquid surfaces do not follow Rayleigh scattering, but they often strongly contribute to the reflectivity of the exoplanet.

In this section we describe a novel approach to the reflectance properties of an exoplanet atmosphere to include effects of a greater atmospheric variety.

We assumed that any non-Rayleigh single-scattering from the atmosphere follows Lambertian scattering. This assumption is discussed below. A planet is better defined by two combined albedo quantities:
  \begin{itemize}
    \item a total geometric albedo \Agtl that represents the total fraction of incoming irradiance reflected by the exoplanet, taking into account all scattering processes,
    \item a Rayleigh single-scattering albedo $\sal$ that represents the fraction of incoming irradiance that is scattered by the atmosphere according to Rayleigh single-scattering alone, from which arises a Rayleigh geometric albedo \Agrl. Only this part of reflected light carries a defined degree of polarization in this model. This albedo can be approximated by the single-scattering albedo of the exoplanet multiplied by the fraction of Rayleigh-scattered emerging light.
  \end{itemize}

These albedo values can be added linearly so that
  \begin{equation}
      \Agtl=\Agll+\Agrl.
  \end{equation}

Following this, we can express the outgoing irradiances using the phased albedo \phl:
  \begin{align}
      \phl_{,Rayleigh}&=\pfl_{Rayleigh} \cdot \Agrl \label{PhasedPoldef}\\
      \phl_{,Total}&=\pfl_{Lambertian} \cdot (\Agtl-\Agrl)
      \notag\\&~~~~+\pfl_{Rayleigh} \cdot \Agrl. \label{PhasedIntdef}
  \end{align}

The total emerging spectral irradiance of the exoplanet $I_{Total,out}$ including thermal emission is
  \begin{equation}\label{intensity}
      I_{Total,out}=E_{Rayleigh,out}+E_{Lambertian,out}+E_{Thermal,out}.
  \end{equation}
Note that this thermal emission is much weaker than the reflected light from the host-star unless the exoplanet is a hot Jupiter and the observing wavelength lies in the infrared. Below, we assume that $E_{Thermal,out}$ is negligible for $\lol\lesssim1\mu m$.

The different Stokes parameters and polarization degree can be easily obtained with
  \begin{align}
      Q/I&=\frac{(\frac{Q}{I_{Rayleigh,out}}) \cdot E_{in} \cdot \phl_{,Rayleigh}}{I_{Total,out}} \label{Qout}\\
      U/I&=\frac{(\frac{U}{I_{Rayleigh,out}}) \cdot E_{in} \cdot \phl_{,Rayleigh}}{I_{Total,out}} \label{Uout}\\
      V/I&=0 \label{VsurI}\\
      \dpl_{Total}&=\frac{\dpl_{Raylgeigh}}{1+\frac{\phl_{,Lambertian}}{\phl_{,Rayleigh}}}, \label{polout}
  \end{align}
where $Q/I_{Rayleigh,out}$ and $U/I_{Rayleigh,out}$ are the Stokes ratios from the Rayleigh-scattering polarization section of this model; they are used to calculate $\dpl_{Raylgeigh}$ using Equation~\ref{polfractioneq}.

According to previous sections, the following complete relation for the exoplanet reflected irradiance, given the host-star surface irradiance, is
  \begin{equation}\label{fullrelstarpla}
      E_{Planet,Distance}=E_{Star,Surface} \cdot (\frac{R_{\ast}}{\rl})^2 \cdot (\frac{R_{Planet}}{d})^2 \cdot \phl,
  \end{equation}
where $R$ are radii. The observer-exoplanet distance is here approximated by the distance observer-star $d$. The contrast ratio \crl between a planet and its host-star is then obtained from Equation~\ref{fullrelstarpla}:
  \begin{equation}\label{ratiostarpla}
      \crl=(\frac{R_{Planet}}{\rl})^2 \cdot \phl.
  \end{equation}

We highlight here that the polarization of the exoplanet $\dpl_{Total}$ and the contrast ratio \crl do not depend on the incoming irradiance from the host-star, they are intrinsic to the planet. However, they are strongly dependent on the observation wavelength \lol through Rayleigh single-scattering and geometric albedo, even though for clarity of notation, $\lambda$ was not explicitly written in the relations.

Thanks to the separation of the contributions from the different scattering processes on an exoplanet, an original model for planetary polarized reflectance has now been set up. It yields the phased albedo \phl, which indicates the fraction of reflected irradiance from an observed planet, as a function of geometric albedo \Agl, phase angle \pal, and Rayleigh single-scattering albedo \sal. While \pal can be easily determined as described in the orbital computations part of the model, the albedo parameters \Agl and \sal depend on factors too numerous to be modeled here: they therefore need prior computation. If \Agl and \sal are obtained separately, the albedo and phase functions previously defined are successful in describing the emerging irradiance that is reflected by a planet. Its most accurate results are obtained for wavelengths shorter than approximately $1\mu m$ unless thermal emission and Mie scattering are also added to the model.

We assumed a polarized-reflectance model that incorporates Rayleigh and Lambertian scattering as two linear scattering phenomena. The approximation made is that any scattering that does not relate to Rayleigh single-scattering is assumed to be Lambertian. As a consequence, a beam of light undergoing multiple-Rayleigh scattering or a combination of Lambertian and Rayleigh scattering is assumed to be multiple Lambertian scattered. This approximation mainly affects the angle at which a photon emerges from the exoplanet and in turn the phase-function of integrated emerging light. In our model, this integrated phase function is a linear combination of Rayleigh and Lambertian phase functions. Adding these second-order scattering phenomena to the computation of the integrated phase function adds terms to its computation: a multiple Rayleigh phase function term and a coupled Rayleigh-Lambertian phase function term. We assumed that these second-order phase functions are similar to the Rayleigh 
and Lambertian phase functions, which differ only slightly (see Figure~\ref{phasefunctions}). Furthermore, we assumed that the weight of these additional phase functions in the computation of the integrated phase function is smaller than Rayleigh single-scattering and Lambertian single or multiple scatterings.
Note also that the lower the albedo values, the less likely multiple-scattering becomes and the smaller the errors induced by this approximation.

The results were benched-marked against several other models providing photopolarimetric curves as a function of one or several orbital parameters, such as models developed by~\citet{2009A&A...504..259B},~\citet{2012ApJ...747...25M},~\citet{2010A&A...512A..59F}, and ~\citet{2010ApJ...723.1168Z}. The locations of the polarization peaks and minima were reproduced with very good agreement; they mostly depend on the phase functions and the processing of orbital parameters. The intensity of the polarization peaks were reproduced with good accuracy; they mostly depended on the fine-tuning of the two albedo values. More specifically, the shift of the polarization peak to phase angles greater than $90\degree$ reported in the last bench-mark reference was also observed.


\subsection{Model completeness}

The calculation code that implements this model takes up to 11 input parameters and N orbital positions for which the calculations are performed. It generates N-element vectors for up to 11 output parameters, listed in Table~\ref{output_param}. Table~\ref{compl_param} shows the mapping between the input and output parameters, where the input parameters are listed as column headers and output parameters are row headers of Table~\ref{compl_param}.

\begin{table}[h!]
\centering
\begin{tabular}{ | l | l | l | }
\hline
  Output & Name & Unit \\ \hline\hline
  \pal & Phase angle & degrees  \\ \hline
  \frl & Period fraction since last periapsis & \% \\ \hline
  \nl & True anomaly & degrees \\ \hline
  \rl & Star-planet distance & AU \\ \hline
  \sepl & Angular separation with star & arcsec \\ \hline
  $date$ & Julian Date & JD \\ \hline
  \crl & Contrast ratio with star & $\varnothing$ \\ \hline
  \dpl & Polarization degree & \% \\ \hline
  \apl & Polarization angle & degrees \\ \hline
  $\varphi_{North}$ & Angle with the north & degrees \\ \hline
  $\upsilon_{orb}$& Apparent orbital velocity & mas/hour \\ \hline
\end{tabular}
  \caption{Output parameters from the calculation code. Their processing depends on the availability of input parameters, refer to Table~\ref{compl_param}. ``mas'' stands for milli-arcsecond.}
  \label{output_param}
\end{table}

\begin{table}[h!]
\centering
\begin{tabular}{ | l | l | l | l | l | l | l | l | l | }
\hline
   & \el, \il, \wl & \al & $d$ & \Tl & \tpl & $R_{P}$, \Agl & \sal & \Ol \\ \hline\hline
  \pal, \frl, \nl & x &  &  &  &  &  &  & \\
  \rl$^{1}$, \sepl$^{1}$ &  &  &  &  &  &  &  &  \\ \hline
  $date~^{1}$ & x &  &  & x &  &  &  &  \\ \hline
  $date~^{2}$ & x &  &  & x & x &  &  &  \\ \hline
  \rl$^{2}$, \sepl$^{2}$ & x & x & x &  &  &  &  &  \\ \hline
  \crl$^{1}$ & x &  &  &  &  & x &  &  \\ \hline
  \crl$^{2}$ & x & x &  &  &  & x &  &  \\ \hline
  \dpl, \apl$^{1}$ & x &  &  &  &  & x & x &  \\ \hline
  \apl$^{2}$ & x &  &  &  &  & x & x & x \\ \hline
  $\varphi_{North}$ & x &  &  &  &  &  &  & x \\ \hline
  $\upsilon_{orb}$ & x & x & x & x &  &  &  & x \\ \hline
\end{tabular}
\flushleft
\begin{tabular}{ l l }
      \begin{small}$^{1}$ relative values\end{small} \\
      \begin{small}$^{2}$ absolute values \end{small}                           
\end{tabular}
  \caption{Mapping between the input (columns) and output (lines) parameters of the model. An ``x'' indicates that the given input parameter(s) is required for computing that output parameter(s). Example: calculating the relative angular separation (line 4) and the absolute contrast ratio (line 7) only requires six input parameters: \el, \il, \wl, \al, $R_{Planet}$, and \Agl.}
  \label{compl_param}
\end{table}

In the way the code is written, the only mandatory input is the eccentricity, the inclination, and the argument at periapsis. When compared with exoplanetary detections reported in the literature, these three orbital parameters are rarely known with any precision. This is currently the case for 132 exoplanets out of the 1790 ($\approx$7.4\%, source: ~\url{http://exoplanet.eu/}). However, where more complete data exist, reasonably good estimates can be made.


\section{Results}

\subsection{SPOC diagrams}
An important tool introduced here is the Separation-POlarization-Contrast (\diag) which presents all the useful information to enable an observer to evaluate whether a planet target is observable at given instrumental limits and how this signal will evolve with orbit and time. Perhaps most importantly, in the event of a detection, \diags provide a powerful mechanism to constrain exoplanet properties given observational imaging data.

\begin{figure}[h!]
  \centering
  \includegraphics[width=\hsize]{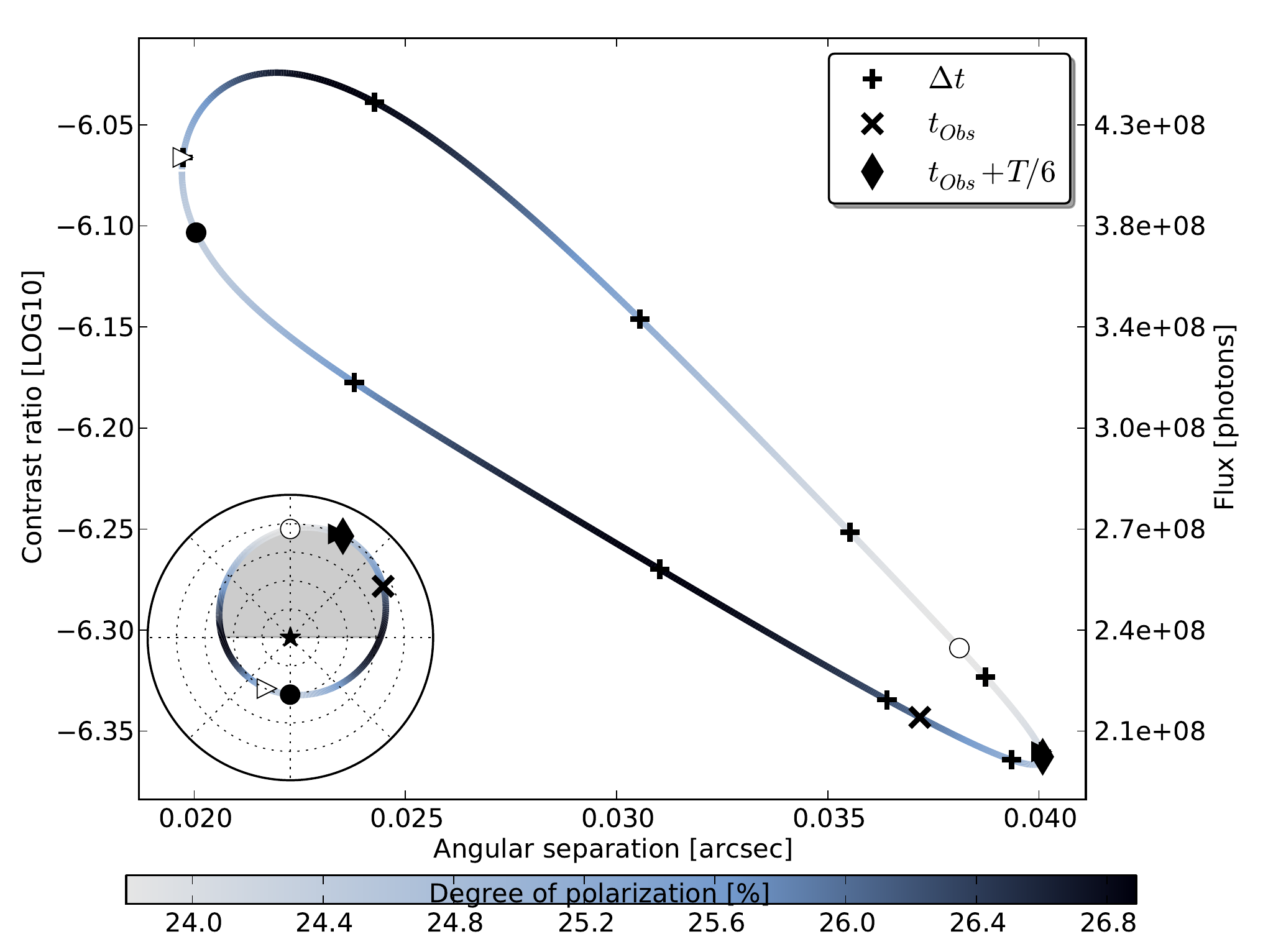}
    \caption{\diag for $Cen_{\alpha,0.34}$ with \il=15\degree, see Table~\ref{Alforbparam}, assumed to be Venus-like regarding its reflectance parameters. It represents the polarization values as gray scale, plotted against the angular separation and the contrast ratio axes. The exoplanet's orbit is therefore displayed in a separation-contrast phase diagram. The wavelength is the V-band. The ``$\textbf{x}$'' indicates the position of the exoplanet at an arbitrary observation time. The `diamond indicates the position of the planet one sixth of the orbital period \Tl later (giving the forward direction of time). The 10 plus signs indicate the evolution of the planet along the curve; they are linearly spaced in time.
    The empty and filled triangles are periapsis and apoapsis: $r_{min}=0.026AU$ and $r_{max}=0.054AU$. The empty and filled circles are minimum and maximum phase angle: $\pal_{min}=75\degree$ and $\pal_{max}=105\degree$. The bottom left panel shows the orbit as seen by the observer, where the shaded area corresponds to the portion of the orbit that is behind the plane of the paper. The dashed lined corresponds to the limit of the same instrument sensitivities as in the preceding example. The right-hand side y-axis is given in photons per VLT collecting area per hour integrated over V-band. Two successive plus signs correspond to 7h46min.}
  \label{examplefourD}
  \end{figure}

Figure~\ref{examplefourD} shows an example of a \diag applied to \object{Alpha Centaury Bb}, an Earth-mass planet discovered in 2012~\citet{2012Natur.491..207D} that was later debated~\citet{2013ApJ...770..133H}. Its probable orbital parameters are shown in Table~\ref{Alforbparam}. The exoplanet was discovered with the radial velocities method, hence its inclination and argument of ascending node are unknown. For \il, values in [15, 35, 60, 85]\degree were explored, whereas \Ol was neglected because it does not change the shape of the orbit: it only defines a rotation of the orbit locus as seen by an observer along the line of sight (i.e., it is only useful for projecting the exoplanet location around the star onto RA-DEC axes). Its radius was calculated from its mass$\cdot sin(\il)$ = 0.0036 $\cdot M_{Jupiter}$ assuming the same density as Earth, which leads to M = [4.42, 2.00, 1.32, 1.15] $M_{Earth}$ and R = [1.64, 1.26, 1.10, 1.05] $R_{Earth}$ for the four previous \il. In this example, the light from \
object{Alpha 
Centaury A} received by the exoplanet is neglected because it is roughly $2.10^5$ 
fainter than that from \object{Alpha Centaury B}. Reflectance parameters were chosen to be similar to those of a Venus-like planet (\Agl=0.67, \wl=0.85) in V-band.

  \begin{table}[h!]
    \begin{center}
      \begin{tabular}{ | l | l | l | l | l | l | } \hline
      Set & $\Tl~[day]$ & $\wl~[\degree]$ & $\al~[AU]$ & $\el$ & $\tpl~[JD]$ \\ \hline \hline
      $Cen_{\alpha,0}$ & 3.2357 & 0 & 0.04 & 0 & 55280.17 \\ \hline
      $Cen_{\alpha,0.34}$ & 3.2357 & 246 & 0.04 & 0.34 & 55282.53 \\ \hline
      \end{tabular}
      \caption{\label{Alforbparam}Two sets of probable orbital parameters for \object{Alpha Centaury Bb}, see section 9 of the supplementary information of~\citet{2012Natur.491..207D}. The inclination is choosen from [15, 35, 60, 85]\degree, which leads to M = [4.42, 2.00, 1.32, 1.15] $M_{Earth}$ and R = [1.64, 1.26, 1.10, 1.05] $R_{Earth}$ respectively (assuming Earth density).}
    \end{center}
  \end{table}

In Figure~\ref{examplefourD}, $Cen_{\alpha,0.34}$ is shown with \il=15\degree. This figure displays the polarization encoded with color, while the two axes plot the separation against contrast ratio; finally, a set of different markers give temporal information. The first striking feature of the plot is that the exoplanet does not run a ``back and forth'' locus in the separation-contrast ratio phase diagram. This is because the apparent symmetry of the orbit is broken. When the orbital parameters \il and \el are non-zero and \wl is different from [0,90,180,270]$\degree$, the geometry of the orbit as it appears to the observer has no symmetry. In other words, the star is not at a focal point of the apparent elliptical locus of the orbit. Note that the semi-major axis \al does not play a role in the shape of the orbit; it acts as a simple scaling factor.
The photon flux on the right-hand side y-axis is given in photons per collecting-area per hour, integrated over V-band. It was calculated assuming that \object{Alpha Centaury B} is a black-body, $T_{eff}=5214$ and $M_v=1.33$. The collecting area is 51.7$m^2$, which corresponds to a diameter of 8.2 meter  with a central obstruction of 2\% (same as VLT telescopes).

Figure~\ref{examplefourD} shows that the lowest and highest contrast ratios are reached close to the maximum and minimum phase angle (black and white disks). This corresponds to the inferior and superior conjunction of the exoplanet with its star. We note that the periapsis and apoapsis are not reached at these phase angle extrema because of the non-null value of \wl, which re-orients the orbit with respect to its host star. Polarization reaches two maxima near \pal=90\degree (the limit between the shaded and the bright area in the bottom left subplot of the same figure). This is linked to the fact that polarization is observed from Rayleigh scattering, which has a peak polarization for 90\degree. However, in the case of non-entirely Rayleigh planets, the peak polarization occurs at slightly different phase angles.

\begin{figure}[h!]
  \centering
  \includegraphics[width=\hsize]{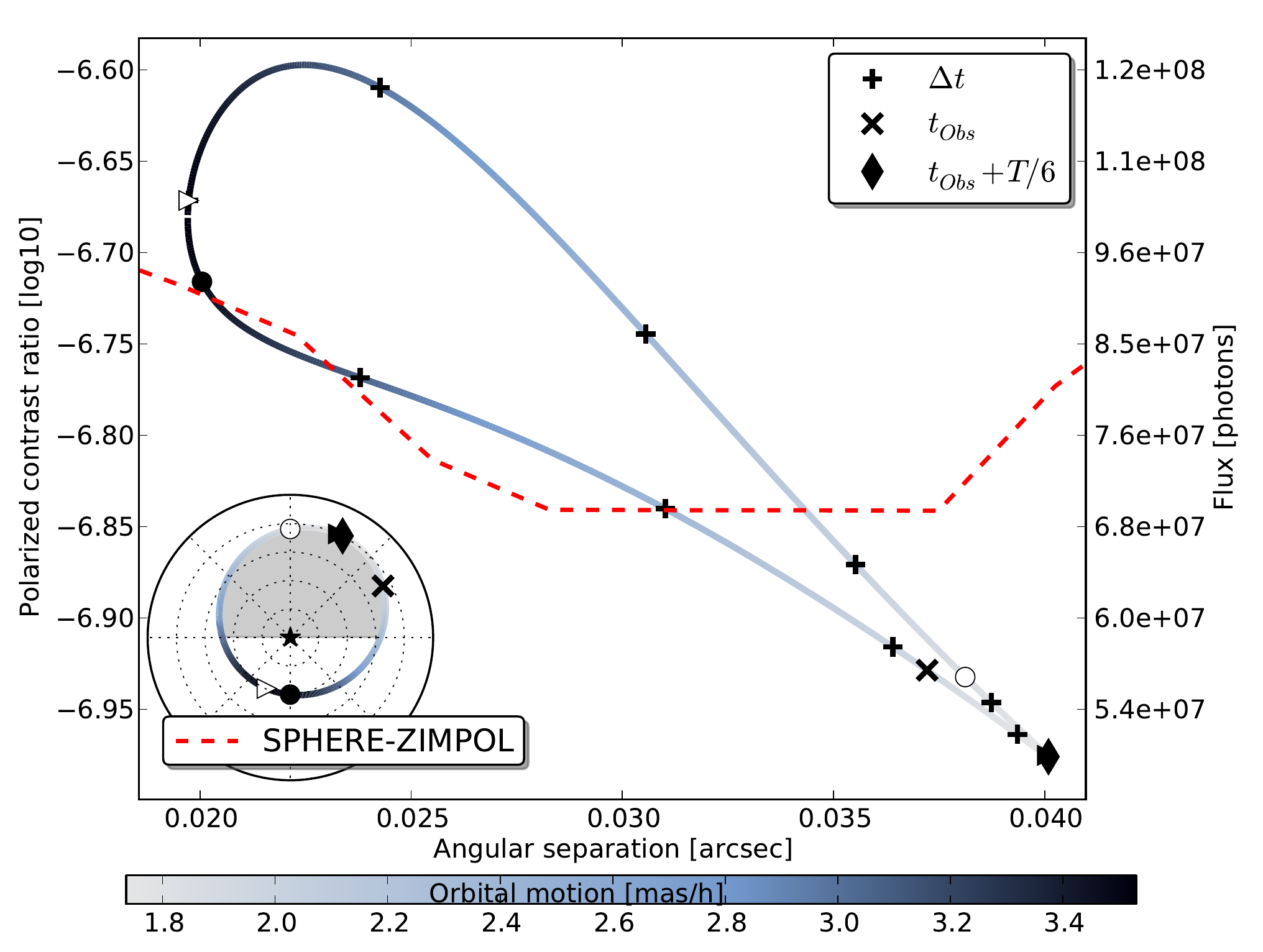}
    \caption{Same caption as Figure~\ref{examplefourD} except that it represents the apparent orbital motion in mas per hour as gray scale, plotted against the angular separation and the polarized contrast ratio axes. The red dotted line corresponds to the SPHERE-ZIMPOL performance for 1 hour exposure (observability domain being above the line).}
  \label{examplefourDpol}
  \end{figure}

An even more interesting diagram is shown in Figure~\ref{examplefourDpol}. It represents the apparent orbital motion in mas per hour plotted against the angular separation as x-axis and the polarized contrast ratio as y-axis. The apparent orbital motion is found to be an important factor to take into account here because the period of \object{Alpha Centaury Bb} is only 3.24 days: it reaches 1.8 to 3.5 mas per hour, which limits the observer to a few hours of exposures, depending on the plate scale of its detector. The photon flux is extremely high as a result of the brightness of the star (1.3mag in V) and the large bandwidth of the V filter ($550\pm44$nm).
The SPHERE-ZIMPOL performance curve for 1h exposure time using a V-filter with the double-difference polarization calibration was added to the plot (see Fig. 4 in ~\citet{2014SPIE.9147E..3WR}). It shows that this instrument could observe \object{Alpha Centaury Bb} during $\approx35\%$ of its orbit if \il=15\degree. Longer integrations could increase the sensitivity at the expense of larger apparent orbital motion on the detector, which will significantly limit the angular differential imaging (ADI) processing capability.

\begin{figure}[h!]
  \centering
  \includegraphics[width=\hsize]{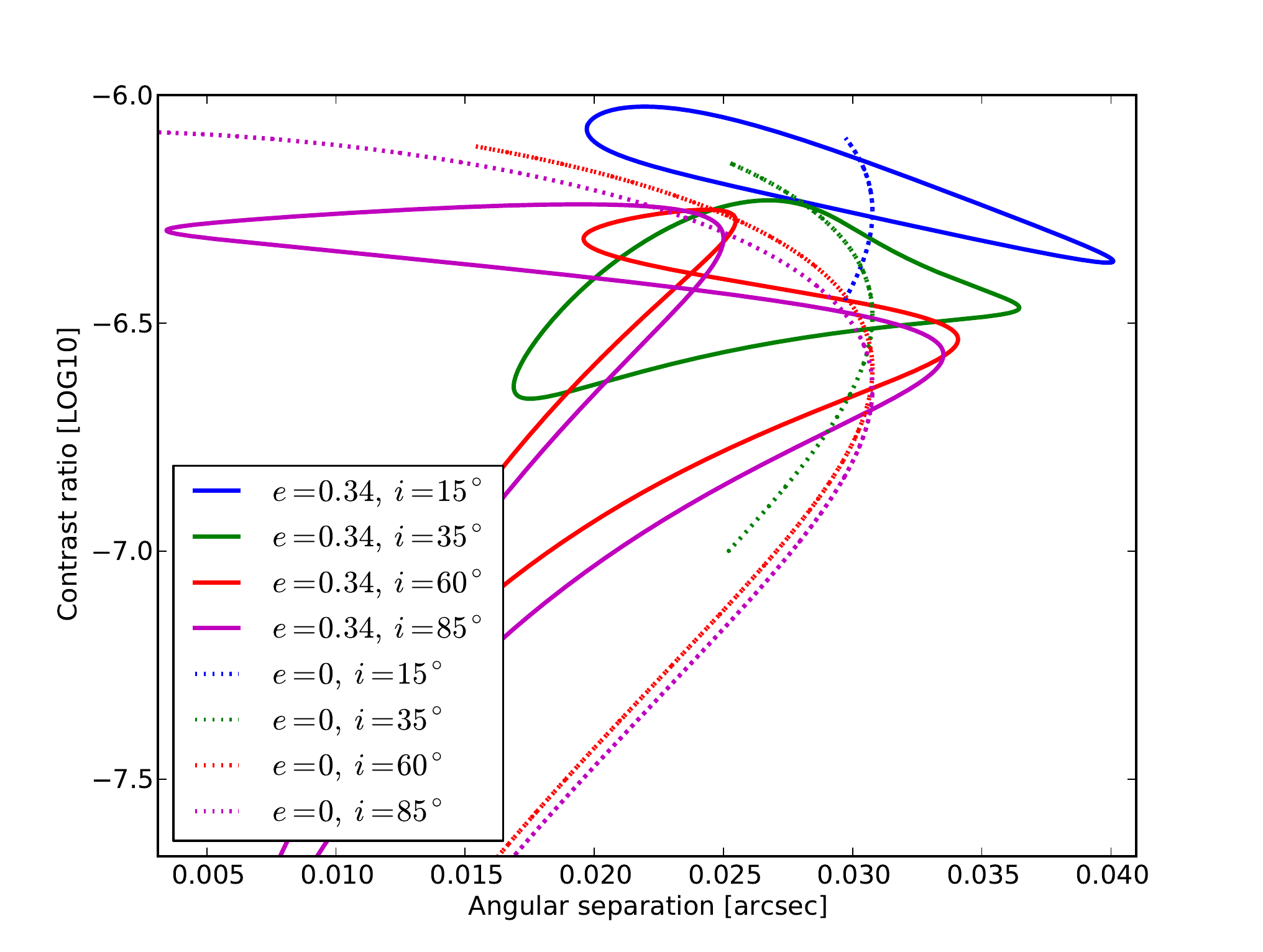}
    \caption{Contrast ratio versus angular separation plots for both orbital parameters of \object{Alpha Centaury Bb} as described in Figure~\ref{examplefourD} and for inclinations in [15, 35, 60, 85]\degree. Dotted lines correspond to \el=0 and \wl=0\degree and full lines to \el=0.34 and \wl=246\degree. A color version is available in the online journal.}
  \label{incli_alfcen}
  \end{figure}

Figure~\ref{incli_alfcen} shows \diags for the two sets of orbital parameters $Cen_{\alpha,0.34}$ and $Cen_{\alpha,0}$, and for all four \il.

This figure shows that for higher orbital inclinations, the radius inferred from the orbital velocity profile decreases, which leads to a smaller reflecting surface and consequently, a lower contrast ratio. The effect of eccentricity is clearly seen: curves for $Cen_{\alpha,0}$ all run a ``back and forth locus'' because of their orbital symmetry. Eccentric orbits display an increasingly complex shape as the inclination increases: this is due to the projection of the orbit locus on the sky as an observer would see it.

We refer to Appendix~\ref{app:appendixsevenreg} for more detailed studies of diverse \diags that are based on known exoplanets.


\subsection{Separations and contrast ratios for the solar system}

An interesting illustration of the usefulness of the \diags is provided by our solar system as viewed by an external observer.

  \begin{table*}
  \centering
    \begin{tabular}{|l|l|l|l|l|l|l|}
    \hline
    Planet & \multicolumn{1}{l|}{Diameter to Earth} & \al $[AU]$ & \el & \il $[\degree]$ & \wl $[\degree]$ & \Agl \\ \hline \hline
    Mercury & 0.382 & 0.3870 & 0.205 & 7.00 & 77.45 & 0.142 \\ \hline
    Venus & 0.948 & 0.7233 & 0.006 & 3.39 & 131.53 & 0.67 \\ \hline
    Earth & 1 & 1.000 & 0.016 & 0.00 & 102.94 & 0.367 \\ \hline
    Mars & 0.532 & 1.523 & 0.093 & 1.85 & 336.04 & 0.170 \\ \hline
    Jupiter & 11.2 & 5.203 & 0.048 & 1.30 & 14.75 & 0.52 \\ \hline
    Saturn & 9.44 & 9.537 & 0.054 & 2.48 & 92.43 & 0.47 \\ \hline
    Uranus & 4.00 & 19.19 & 0.047 & 0.76 & 170.96 & 0.51 \\ \hline
    Neptune & 3.88 & 30.06 & 0.008 & 1.76 & 44.97 & 0.41 \\ \hline
    \end{tabular}
  \caption{Planetary and orbital parameters for the planets of the solar system, in the V-Band. Source:~\url{http://nssdc.gsfc.nasa.gov/planetary/factsheet/}. Inclinations is this table are measured from the plane of the ecliptic.}
  \label{ssol_table}
  \end{table*}

\begin{figure*}
  \centering
    \includegraphics[width=17cm]{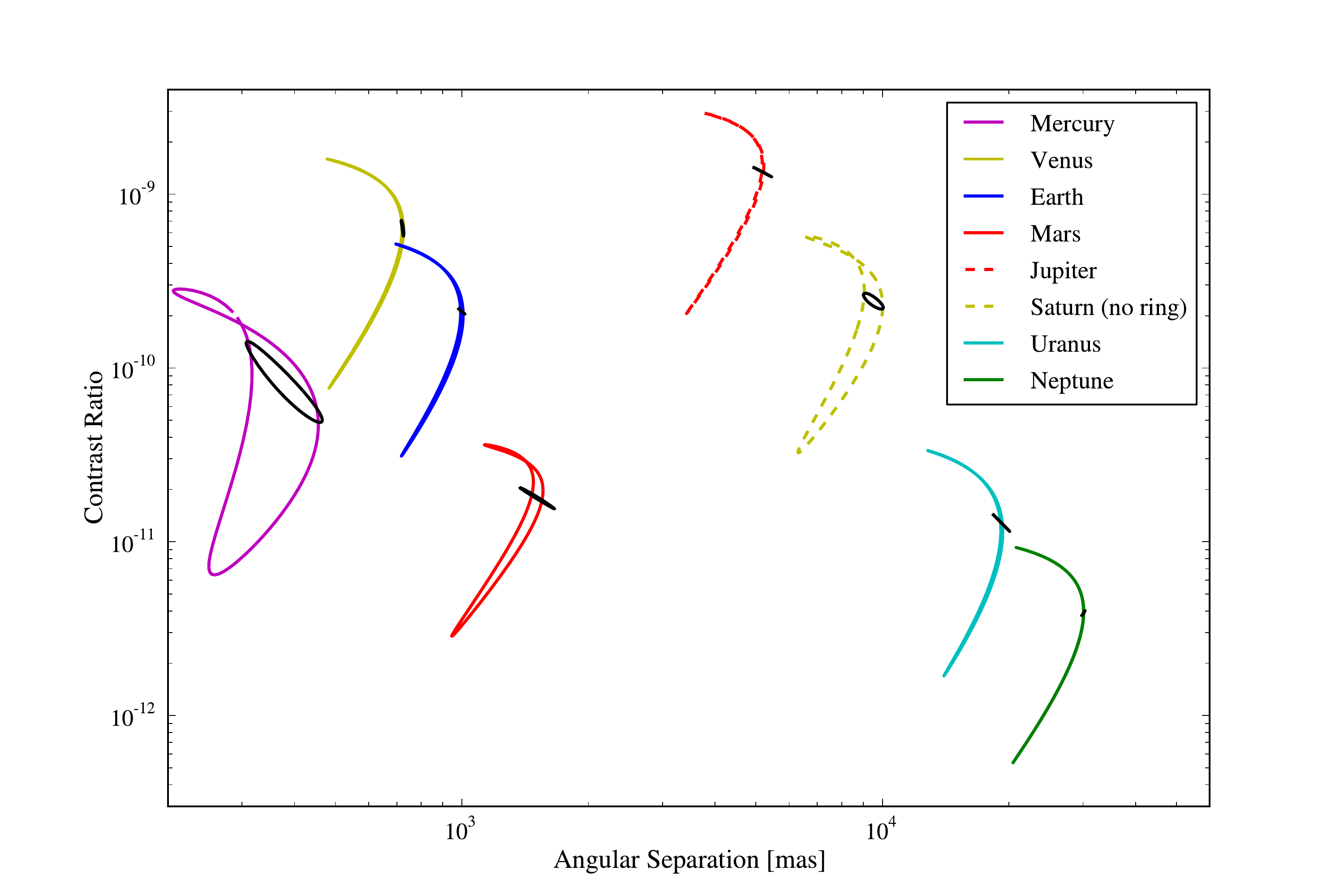}
      \caption{Contrast-separation diagram for the planets of the solar system as seen from 1pc distance in visible light. Small black curves correspond to an inclination of the ecliptic of 0\degree (observer reference frame); colored curves show the solar system at an inclination of 45\degree. Note that both axes have logarithmic scales. A color version is available online.}
      \label{ssol}
  \end{figure*}

Figure~\ref{ssol} gives the contrast-separation diagram for the planets of the solar system, whose parameters are listed in Table ~\ref{ssol_table}. In this figure, polarization is not shown (for clarity). It displays values for two different angles of inclination for the observer, $0$\degree and $45$\degree. The observational wavelength is visible light. Note that no rings were modeled for Saturn; they are anticipated to have a variable effect on both contrast ratio and polarization (see discussion in Sect. 3.3).

Note that the angular separation of Earth to the Sun is $\approx1$arcsec when the ecliptic inclination is $0$\degree, consistent with the definition of a parsec. Earth's average angular separation decreases as the orbit appears more inclined to the observer.

Figure~\ref{ssol} highlights the strong dependency of the contrast ratio and angular separation on inclination, argument at periapsis, and eccentricity. Mercury especially, but also Mars and Saturn do not follow a ``back and forth locus'' because the symmetry of their orbit as it appears to the observer is broken.

Venus is seen to be the second brightest planet in the solar system (for a ringless Saturn). Significantly brighter than Earth, but fainter than Jupiter. A rule of thumb can explain this easily: Venus receives twice as much solar flux as Earth and has an albedo nearly twice as high. Because the diameters of the two planets are close to each other, they present almost the same reflective area (Venus has $\approx$90\% of that of Earth): Venus is therefore on average nearly four times brighter than Earth to an external observer. The same rules can be applied to compare the relative brightness of Jupiter and Venus': Jupiter is only twice as bright as Venus to the observer.
The same rule of thumbs can explain the surprising faintness of Mars, Uranus, and Neptune. They show a relatively small reflective area because of their distance to the Sun, which makes them receive little flux; furthermore, Mars has a significantly low albedo. Note that these comparative values are averaged over the entire orbits. 

\begin{figure}
  \centering
    \includegraphics[width=\hsize]{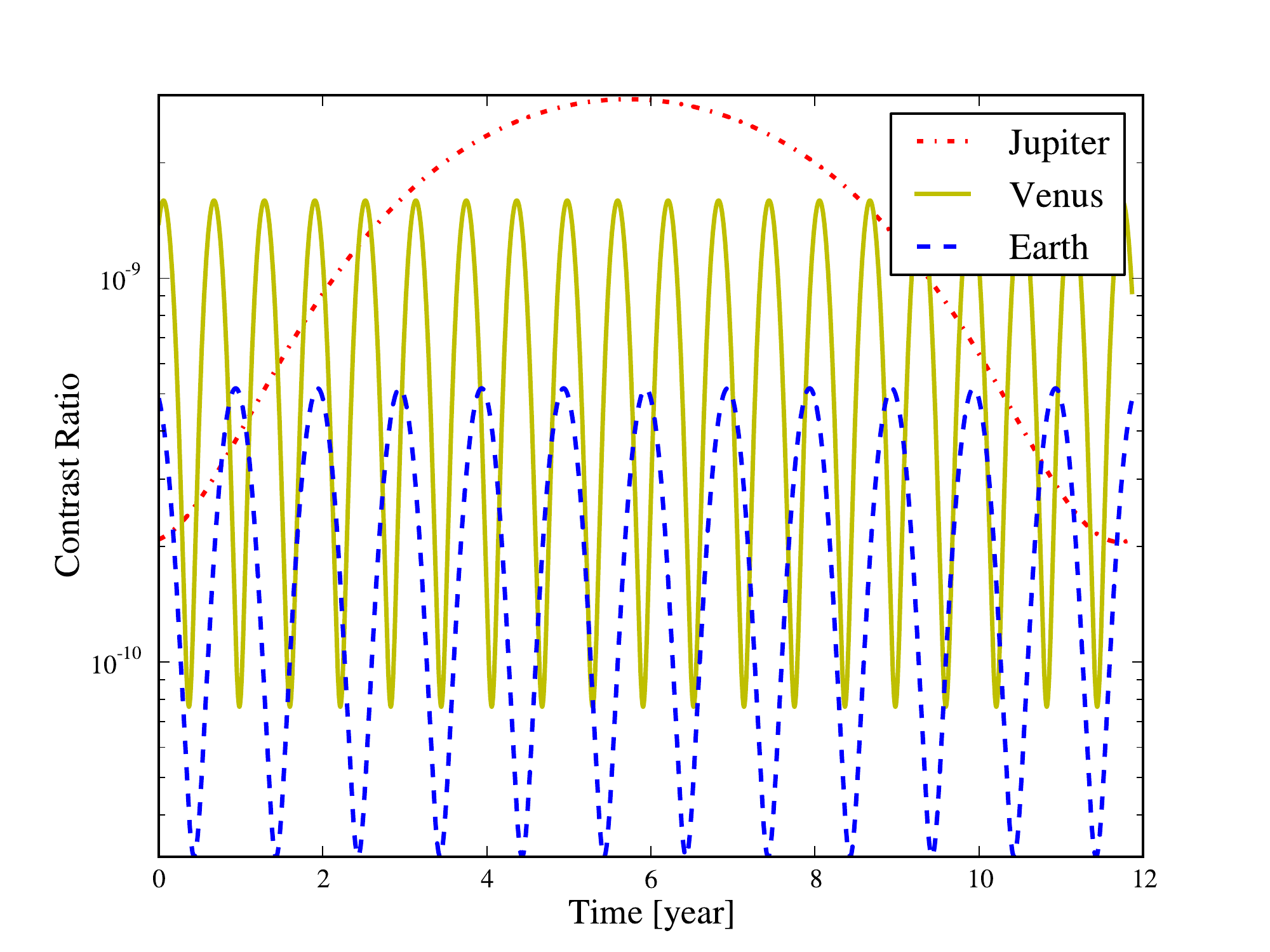}
      \caption{Evolution of the contrast ratio to the Sun for Venus, Jupiter and the Earth, during an entire Jovian revolution (11.9 years).}
      \label{brightest}
  \end{figure}

Finally, this same figure highlights once more that focusing the discussion on averaged contrast ratios and separation is futile: depending on the respective configuration of the planets, Earth might easily become the brightest planet in the solar system at some epochs. As an example, let us consider the case when the observer sees the Solar System with an inclination of 45\degree. Figure~\ref{brightest} shows the contrast ratios of Jupiter, Venus, and Earth as a function of time during a whole Jovian orbit.

It can be calculated that, during the Jovian revolution (11.9 years):
  \begin{itemize}
    \item CR(Earth) > CR(Venus) for 2.55 years (21.53\% of $T_{Jup}$),
    \item CR(Earth) > CR(Jupiter) for 1.04 years (8.82\%),
    \item CR(Venus) > CR(Jupiter) for 3.16 years (26.61\%),
    \item Venus is the brightest of the 3 planets for 3.11 years (26.16\%),
    \item Earth is the brightest of the 3 planets for 4.83 months (3.39\%),
    \item Jupiter is the brightest of the 3 planets for 8.36 years (70.44\%),
  \end{itemize}where CR stands for contrast ratio. These values become 28.28\%, 18.89\%, 32.14\%, 30.30\%, 4.83\%, and 64.86\% respectively when the observer sees the solar system at an inclination of 75\degree. Note also that for an inclination of 45\degree, Mercury is brighter than Jupiter during 1.75\% of $T_{Jup}$. This duration increases to 11.62\% for an inclination of 75\degree.

Despite the large variability in apparent brightness, it remains true that all solar system planets are extremely challenging: Earth's relative magnitude to the Sun is about $\approx+23$mag, at best. Gas giant planets at the snow line such as Jupiter do not always offer dramatic gains in observability, as might naively be thought.

\subsection{Effect of extensive planetary rings}

Rings were not included in this model. To do so would at least double the number of parameters describing a planet (exoplanet obliquity, obliquity at periapsis, ring radii (inner and outer), reflectance parameters, etc). The ring orientation to the observer is critical for determining reflected light from the host-star. Beyond the model of our own solar system, there is little observational data to constrain the speculative range of ring properties. However, their presence around an exoplanet would significantly change the observable properties of planets. Rings can either act as reflectors with a potential polarized enhancement or obstruct of the exoplanet illumination or irradiance, depending on their apparent inclination to the star, the observer and their polarized-reflectance characteristics. Both reflection and obstruction effects are coupled for extensive systems of rings that project shadows onto the exoplanet atmosphere. We refer to~\citet{2005ApJ...618..973D} or~\citet{2006dies.conf..105A} 
for a more detailed discussion of the
impacts of rings on light curves.

\subsection{Distinguishing radius and albedo with polarization}

Knowing both the contrast ratio (or absolute flux) of the planet and its polarization degree, preferably at several orbital positions, allows the observer to distinguish between several atmospheric models of the atmosphere and finally determine its radius and albedo. For a given exoplanet atmosphere and the same unpolarized flux measured, a planet with large radius and low albedo will exhibit a higher polarization degree than a small exoplanet with high albedo. We refer to the relation between peak polarization and single-scattering albedo in~\citet{2012ApJ...747...25M}. Such radius-albedo measurements were carried out in~\citet{2008ApJ...673L..83B} and~\citet{2011ApJ...728L...6B}. 


\section{Conclusions}

\subsection{Model and calculation code}

We have constructed a model to fully describe the emerging radiation field from an exoplanet with given orbital geometry and reflectance parameters. This model aclculates the three main observables that are relevant for direct imaging of an exoplanet: polarization, contrast ratio, and angular separation as a function of date. A key strength of this model is the relatively restricted number of free input parameters despite the complex processes addressed. Only 3 of these input parameters -- eccentricity, inclination, and argument of periapsis -- are mandatory for performing a first assessment of the variability over time of the exoplanet observability.

The calculation code developed is available for public use and collaborative improvement on the python package index~\url{https://pypi.python.org/pypi} under ``exospoc'', together with its documentation. This code is implemented on the~\url{http://exoplanet.eu/} exoplanet database (see~\citet{2011A&A...532A..79S}), where \diags are accessible in the exoplanet information sheets at the link ``Observability Predictor''.

\subsection{SPOC diagrams}

We have introduced a novel tool - the \diag. It highlights the interdependency of polarization, contrast ratio, and angular separation to the host star for an input exoplanet and gives the observer critical information for predicting expected exoplanetary signal from a minimum set of parameters. The complex shape of the \diag curves highlights the fact that in the general case (inclination higher than $\approx$10\degree), the critical observables strongly depend on the geometry of the orbit as it appears to the observer. As a consequence, we stress that restricting consideration to values averaged along the entire orbit for angular separation, contrast ratio, or polarization may be a misleading oversimplification in many applications. An illustration of this is that an external observer would report Earth to be the brightest of the eight planets for a significant fraction of random observations.

The variation of the main observational criteria is critical in timing a direct observation of an exoplanet. This variation mainly relies on the combined effect of the inclination, the eccentricity and the argument of periapsis, hence the shape of the orbit as it appears to the observer. This latter parameter has a surprisingly important role to play in the visibility prediction for a planet in acting to either amplify or cancel the effects of inclination and eccentricity over the contrast ratio and angular separation values over time. The albedo parameters only shifts the lowest and highest values of the contrast ratio; however they do not significantly change the span of its minimum and maximum. The planet radius, semi-major axis and, observer-host-star distance parameters simply scale all observables to higher or lower contrast ratios and angular separations.

The simultaneity of the maxima of three main observational criteria can also be studied with \diags. The simultaneity of contrast ratio and polarization (or of contrast ratio and angular separation) maxima is unlikely (or very unlikely) in the $t_{CR,max}\pm(20$\% of $\Tl)$ temporal window. Therefore, the best-case scenario for detection, maxima of contrast, separation, and polarization, is extremely unlikely.


\begin{acknowledgements}
     We thank Vincent Coudé du Foresto and Sylvestre Lacour for their comments and advice. We acknowledge that this work was partly supported by the French National Agency for Research (ANR-13-JS05-0005-01). We thank the exoplanet.eu team and especially Marco Mancini, who adapted this work to the online database.
\end{acknowledgements}


\bibliography{exospoc_arxiv.bbl}

\clearpage
\appendix

\section{Observability predictor examples}\label{app:appendixsevenreg}

  \begin{figure}[!h]
  \centering
  \includegraphics[width=\hsize]{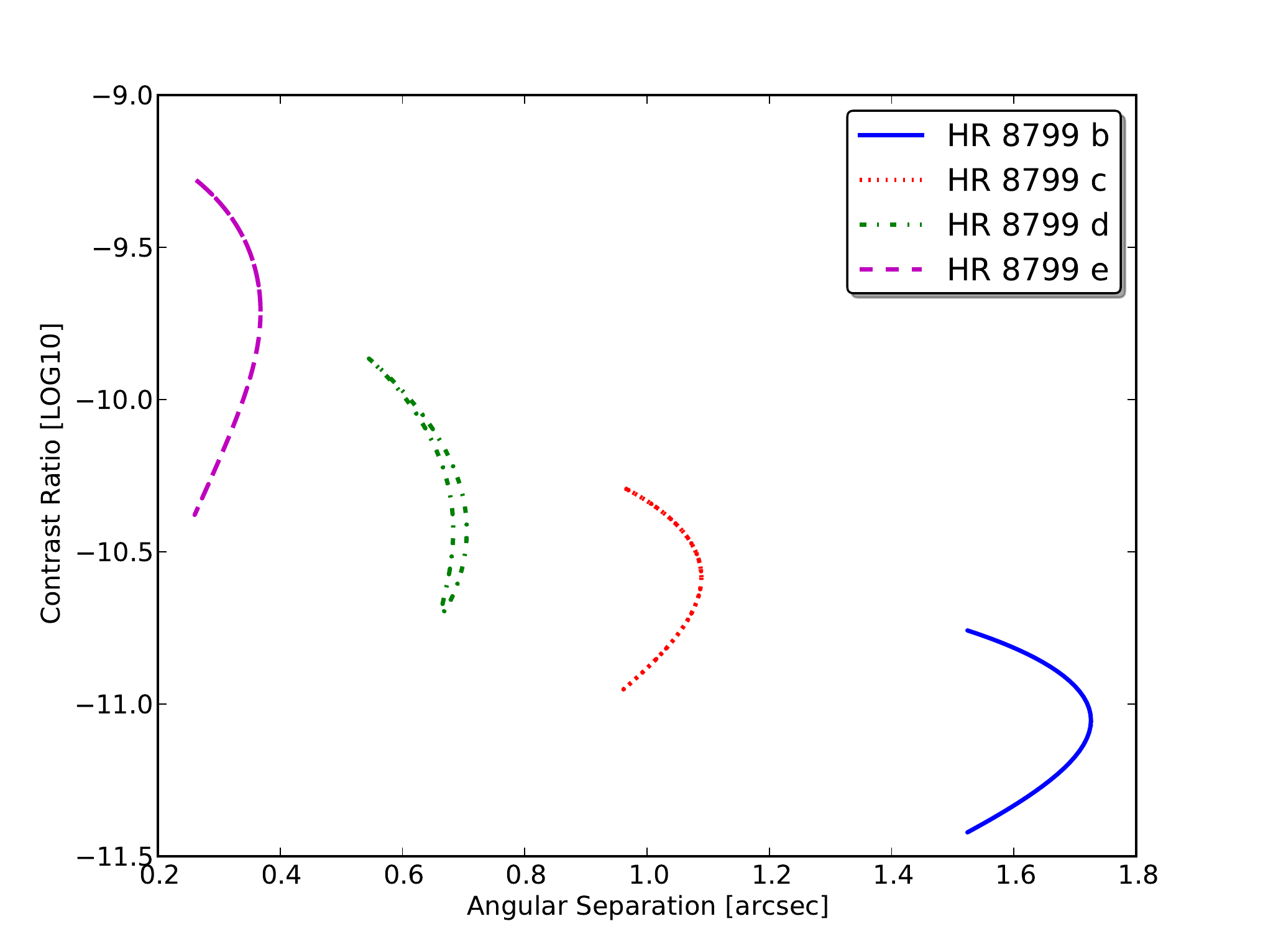}
    \caption{Contrast-separation diagram for V-band reflectance parameters (no polarization information added for clarity reasons) showing the famous \object{HR 8799} exoplanetary system with its four confirmed candidates. The planet e was assumed to be the same size as the three other planets (1.2$R_{Jupiter}$). All four planets but planet d follow a ``back and forth'' locus because of their null eccentricity. Despite their large reflecting area, the contrast ratios are extremely low because of their distance to the host star.}
  \label{region1}
  \end{figure}

  \begin{figure}[!h]
  \centering
  \includegraphics[width=\hsize]{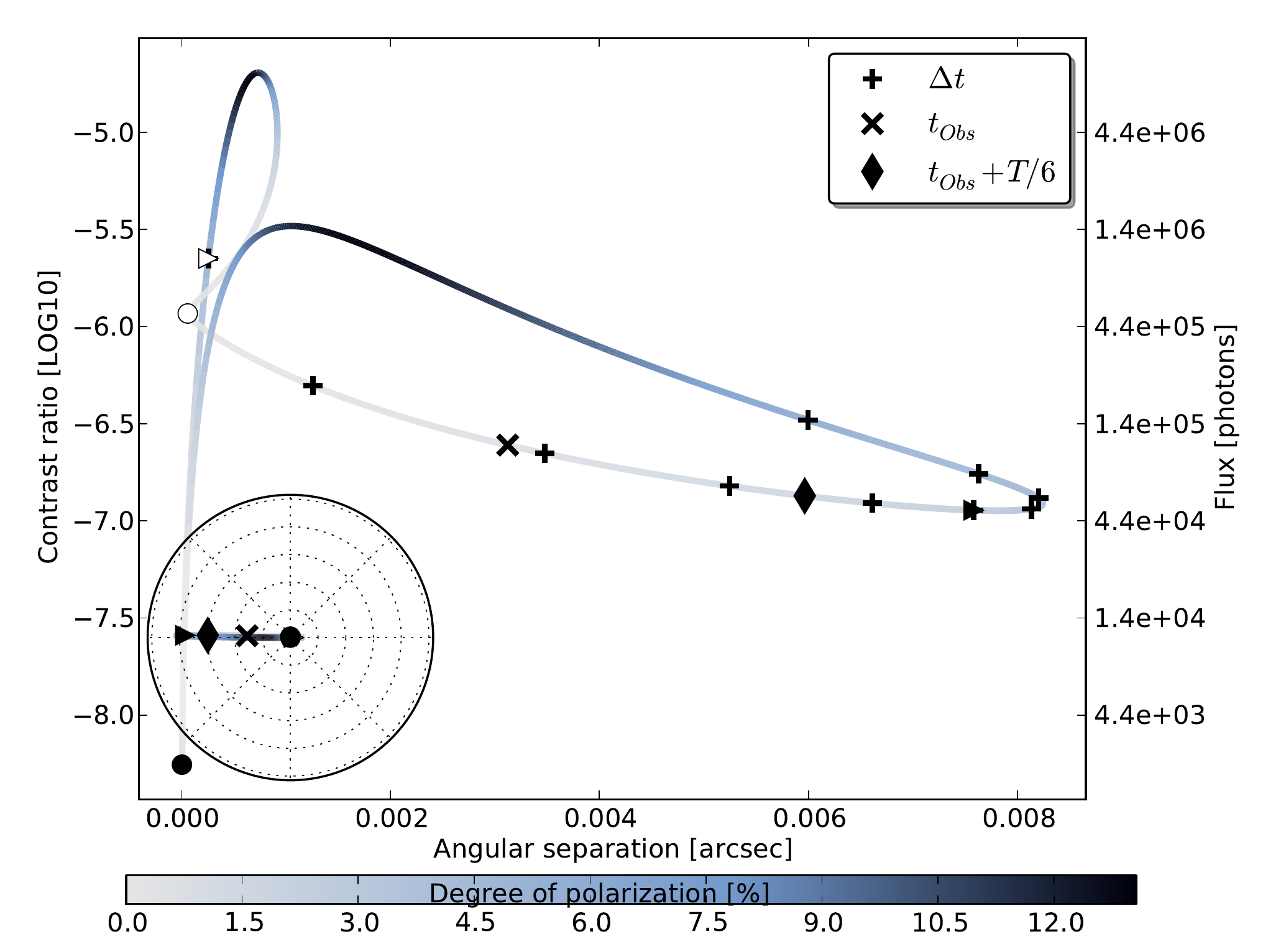}
    \caption{SPOC diagram for \object{HD 80606 b}, for which all orbital parameters are known down to a very high accuracy. The very high eccentricity (0.93) of this transiting planet gives a peculiar phase diagram. The higest contrast ratio is not reached at periapsis because of the combination of \il, \el, and \wl orbital parameters. Photon count is given per hour per VLT collecting area. The reflectance parameters were assumed to be Jupiter-like, in V-band.}
  \label{region2}
  \end{figure}

\end{document}